\documentclass[twocolumn,showpacs,preprintnumbers,amsmath,amssymb,superscriptaddress]{revtex4}

\usepackage{graphicx}
\usepackage{dcolumn}
\usepackage{bm}
\usepackage{braket}
\usepackage{verbatim}
\usepackage{amsmath}
\usepackage{amsfonts}
\usepackage{mathrsfs}
\usepackage{amssymb}
\usepackage{revsymb}
\usepackage{multirow}

\usepackage{color}

\begin{document}

\preprint{APS/123-QED}

\title{Improving Quantum State Estimation with Mutually Unbiased
Bases}
\author{R. B. A. Adamson}
\affiliation{%
Centre for Quantum Information $\&$ Quantum Control and Institute
for Optical Sciences, Dept. of Physics, 60 St. George St.,
University of Toronto, Toronto, ON, Canada, M5S 1A7
}%

\author{A. M. Steinberg}
\affiliation{%
Centre for Quantum Information $\&$ Quantum Control and Institute
for Optical Sciences, Dept. of Physics, 60 St. George St.,
University of Toronto, Toronto, ON, Canada, M5S 1A7
}%

\date{\today}
\begin{abstract}
When used in quantum state estimation, projections onto mutually unbiased
bases have the ability to maximize information extraction per
measurement and to minimize redundancy.
We present the first experimental demonstration of quantum state tomography of
two-qubit polarization states to take advantage of mutually
unbiased bases.  We demonstrate improved state estimation as
compared to standard measurement strategies and discuss how this
can be understood from the structure of the measurements we
use.  We experimentally compared our method to the standard state
estimation method for three different states and observe that
 the infidelity was up to $1.84\pm0.06$ times
lower using our technique than it was using standard state estimation methods.    
\end{abstract}

\maketitle
Quantum state estimation is a central problem in the field
of quantum information, with applications in quantum cryptography,
quantum computing, quantum control, quantum measurement theory and
foundational issues in quantum mechanics.  The practical techniques
used in quantum state estimation such as quantum state
tomography\cite{James2001} have
been pivotal in the recent progress of experimental quantum mechanics.  Many of the major
advances in the field including demonstrations of entanglement
of two\cite{White2002}, three\cite{Resch2005},
and four-photon states\cite{Schmid2007},
quantum logic gate characterization\cite{Obrien2003},
implementation of Shor's algorithm\cite{Lanyon2007}, and cluster state quantum
computing\cite{Walther2005} used quantum state tomography as the
main diagnostic and descriptive tool.  Quantum
state tomography has by now been applied to nearly all candidate
systems proposed for quantum information and computation
including trapped ions\cite{Haffner2004}, spontaneous parametric downconversion
sources\cite{James2001,Mitchell2003,Adamson2007}, atomic ensemble
quantum memories\cite{Choi2008}, atoms trapped in optical lattices\cite{Myrskog2005}, cavity QED systems\cite{Rempe:QTuK4}, quantum dot
sources of entangled photons\cite{Stevenson2006} and superconducting quantum
bits\cite{Steffen2006}.  Improved techniques for quantum state tomography therefore
impact a wide range of applications in experimental physics.   Moreover, because
measurement plays such a central role in quantum mechanics,
quantum state estimation provides one of the best conceptual
tools we have for understanding what quantum states \emph{are}\cite{Leonhardt1996,Gibbons2004}.  

While all tomographic schemes that have been implemented to date for
more than one particle use projection-valued meausurements (PVMs), so
far none has employed the optimal set of PVMs, the set with the
property of being
\emph{mutually unbiased}.  As we will show, mutual unbiasedness removes informational
redundancy among different measurements, a limitation for all
previous multi-particle state estimation strategies.  We
experimentally demonstrate the increase in measurement precision
that can be achieved by employing these special sets of measurements.

Quantum state tomography on qubits
involves the measurement of some linearly-independent,
informationally-complete or
over-complete set of expectation values.  A reconstruction based
on linear inversion\cite{MikeandIke}, maximum-likelihood fitting\cite{James2001} or an
appropriate cost function\cite{Blume-Kohout2010} is then used to calculate the best-fit
density matrix for the data set.


All two-qubit quantum state tomography implementations to date
have constructed a complete, linearly independent set of
projectors from pairwise combinations of eigenstates of the
Pauli operators\cite{MikeandIke}.  Initial implementations
employed $16$ projectors\cite{James2001}, the minimum number required to satisfy
linear-independence and completeness requirements.  Later it was
observed that an improved estimate of
the density matrix could be obtained by performing tomography
with projections onto all 36 tensor
products of Pauli eigenstates\cite{Altepeter2005,deBurgh2008}.
These 36 projectors can be arranged into nine
bases of four orthogonal projectors as shown in the left column of
Table \ref{Table1}.  We will refer to this tomography strategy
as \emph{standard separable quantum state tomography} (SSQST). 

On the face of it, this set of 36 projectors appears unbiased.
Certainly no particular basis or direction is preferred over any
other. If, however, one looks at pairs of bases, then one
notices that some bases share
eigenstates of a particular Pauli operator while others have no
eigenstates in common.  

The overlap among projectors from different bases can be
measured using the Hilbert-Schmidt overlap\cite{MikeandIke}.  Projectors from the first and
second bases of the left side of Table \ref{Table1} that share an eigenstate for
the first qubit have an overlap of $0.5$ whereas those that are
orthogonal in the first qubit have an overlap of zero.  In
contrast, bases that differ in Pauli operators for both qubits have
an overlap of $0.25$ for all pairs of projectors.  

This inequivalence between pairs of bases constitutes a
bias in the measurement scheme.  This bias, which will
occur for any complete set of separable projectors, creates 
redundancy and limits the efficiency with
which new information about the state can be collected.  This is
because in schemes that only contain separable
measurements, correlations can only be
observed in one basis at a time.  In contrast, schemes
that employ joint or entangling measurements are capable of probing
correlations in multiple single-qubit bases at once.  For
example, a singlet-state projection onto
$\frac{1}{\sqrt{2}}\left(\ket{HV}-\ket{VH}\right)$ simultaneously
probes anti-correlation in all bases at once while a measurement
of $\ket{HV}$ only determines the degree of correlation in
$\sigma_z\otimes\sigma_z$, but provides no information about
correlation in $\sigma_y\otimes\sigma_y$ or $\sigma_x\otimes\sigma_x$.

As a figure of merit to gauge the accuracy of an estimation technique
we use the \emph{infidelity}, which characterizes the distance between two
density matrices $\sigma$ and $\rho$.  The infidelity is defined as
$1-F$ where $F$ is the
fidelity\cite{Jozsa1994},  
\begin{align}
F=\left(\rm{Tr}\sqrt{\sqrt{\sigma} \rho \sqrt{\sigma}}\right)^2.
\end{align}
While there are other figures of merit that one could adopt,
the fidelity has some appealing operational and analytic properties
that make it well-suited to the task.  A more detailed
discussion of different figures of merit can be found in \cite{deBurgh2008}.
\begin{table}[t]
\begin{tabular}{|c||p{5cm}|}
\hline
\textbf{SSQST} & \textbf{MUB QST}\\
\hline\hline
 $HH$, $HV$, $VH$, $VV$ & $HH$, $HV$, $VH$, $VV$\\\hline
 $HD$, $HA$, $VD$, $VA$ & $RD$, $RA$, $LD$, $LA$\\\hline
 $HR$, $HL$, $VR$, $VL$ & $DR$, $DL$, $AR$, $AL$\\\hline
 $DH$, $DV$, $AH$, $AV$ & $\frac{1}{\sqrt{2}}\left(RL+iLR\right)$,$\frac{1}{\sqrt{2}}\left(RL-iLR\right)$,\\\cline{1-1}
 $DD$, $DA$, $AD$, $AA$ & $\frac{1}{\sqrt{2}}\left(RR+iLL\right)$,$\frac{1}{\sqrt{2}}\left(RR-iLL\right)$\\\hline
 $DR$, $DL$, $AR$, $AL$ & $\frac{1}{\sqrt{2}}\left(RV+iLH\right)$,$\frac{1}{\sqrt{2}}\left(RV-iLH\right)$,\\\cline{1-1}
 $RH$, $RV$, $LH$, $LV$ & $\frac{1}{\sqrt{2}}\left(RH+iLV\right)$,$\frac{1}{\sqrt{2}}\left(RH-iLV\right)$\\\cline{1-1}
 $RD$, $RA$, $LD$, $LA$ & \\\cline{1-1}
 $RR$, $RL$, $LR$, $LL$ & \\\hline
\hline
\end{tabular}
\caption{The measurement bases used in standard separable quantum state
  tomography (SSQST) and mutually unbiased basis quantum state
  tomography (MUB QST).}
\label{Table1}
\end{table}

\begin{figure}{t}
\resizebox{\columnwidth}{!}{\includegraphics{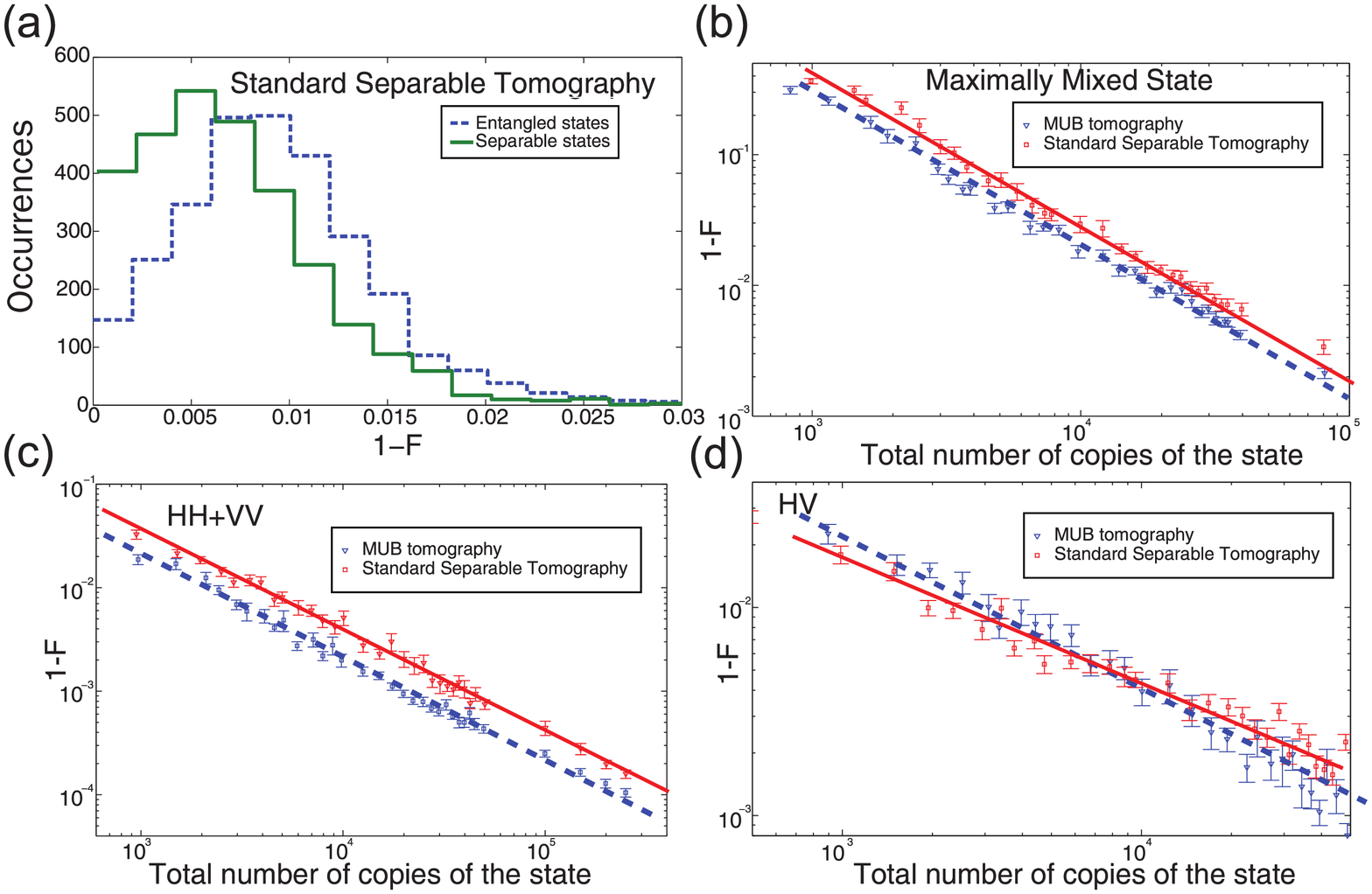}}
  \caption{(a) Histogram of infidelity for $3000$ randomly
    selected entangled states and $3000$ separable states for
    which SSQST was simulated.  The Monte-Carlo simulation
    used $18,000$ total copies for each random state.  (b)
    Comparison of experimental MUB tomography and SSQST for the maximally
    mixed state.  The average value of the ratio of
     infidelity using the SSQST-estimated density
    matrix to that using MUB tomography
    was $1.49\pm0.05$. (c)  Comparison of experimental MUB tomography and SSQST for the state
    $\frac{1}{\sqrt{2}}\left(\ket{HH}+\ket{VV}\right)$.  The
    average ratio of infidelity for the two methods was $1.84\pm0.06$ (d)
    Comparison of experimental MUB tomography and SSQST for $\ket{HV}$.  The
    infidelity ratio for
    the two tomography methods was $1.09\pm0.4$.  In (b) through (d)
    the solid line represents a fit to the SSQST data and the dotted
    line to the MUB QST data.  The apparent crossing of the MUBs and
    SSQST lines in (d) is due to numerical errors at
    high $N$.}
\label{data}
\end{figure}

By this measure, SSQST will generally produce better estimates of
separable states than of entangled states.  This can be observed in
the Monte-Carlo generated data in figure \ref{data}(a) where the
infidelity estimate is plotted as a
histogram over randomly selected maximally-entangled and separable
states.  The states were selected randomly over the Haar measure
induced by local unitary transformations on the states\cite{Zyczkowski1994} and the
estimate was obtained by performing maximum-likelihood fitting
on a simulated data set with on average $18,000$ copies of the
state.  On average, the infidelity is significantly lower
for separable states than for maximally-entangled states.  The median values
of the infidelity for separable and maximally-entangled states are
$0.0054\pm0.0001$ and $0.0091\pm0.0002$ respectivley.  

If the class of measurement bases used in
tomography is augmented to include entangled bases then this limitation of
SSQST can be overcome.  Indeed it is then
possible to achieve \emph{optimal} projective quantum state tomography,
that is to say quantum state tomography with no informational
redundancy.  This can be achieved by taking advantage of mutually unbiased
bases.

\begin{figure}{t}
\resizebox{\columnwidth}{!}{\includegraphics{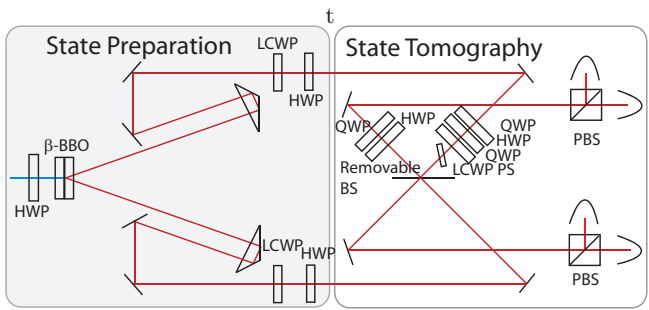}}
  \caption{Experimental apparatus for MUB state tomography.  A
    non-linear crystal (BBO) is pumped to produce pairs of SPDC
    photons.  Liquid crystal waveplates (LCWP), and half-waveplates (HWP)
    rotate the state.  Entangling measurements are made using
    two-photon interference at a removable non-polarizing beamsplitter
    (BS), and these measurements can be rotated with HWPs,
    quarter waveplates (QWPs) and LCWPs to generate all
    necessary projections.  For separable measurements the BS is
    removed and ordinary polarization analysis performed at the
    detectors using the polarizing beamsplitters (PBS), HWPs and QWPs.}
\label{apparatus}
\end{figure}
Mutually unbiased bases (MUBs), first introduced in the context of
quantum state estimation by Wootters and Fields\cite{Wootters1989}, have
the property that all inner products between projectors of different bases labeled $\alpha$ and $\beta$
are equal.  As a consequence,
\begin{equation}
\rm{Tr}\left[P_{\alpha,\gamma}P_{\beta,\delta}\right]=1/D,
\label{equation2}
\end{equation}
whenever $\alpha \neq \beta$, where $\gamma$ and $\delta$ label the PVM
elements for each basis and $D$ is the dimensionality of the Hilbert
space.  The
minimal number of MUBs needed for informational completeness is
$D+1$ since each basis provides $D-1$ independent parameters
plus a normalization, and $(D-1)(D+1)=D^2-1$ is the number of
free parameters in the density matrix.  As it turns out, $D+1$
MUBs are always informationally complete when they exist and
$D+1$ is the maximum number of MUBs that \emph{can} exist.  For qubits
$D=2^N$, and so the $2^N-1$ MUBs bases required for complete
tomography is considerably smaller than the $3^N$ Pauli bases used in
conventional separable tomography.  This can result in a considerable
reduction in experiment time if significant resources are required
to change measurement bases.   

MUBs
can be shown to exist whenever $D$ is the power of a prime, as is the case for all
multi-qubit systems\cite{Wootters1989}.  The existence of a complete set of MUBs in
cases where the dimensionality is not a
power of a prime is an unresolved
problem, although it is widely believed that they do not generally exist\cite{Saniga2004}.

The property stated in equation \ref{equation2}
can be thought of as expressing a complete lack of redundancy
among measurements.  After measuring the projections in one
basis, the probability distribution of possible outcomes in the next
basis is uniform.  In other words, \emph{nothing} is known about
the outcomes of future measurements from previous ones.  More
formally, it has
been shown that MUBs allow the maximum
reduction in the Shannon entropy per
measurement\cite{Wootters1989} averaged over all states.  

This advantage of MUBs is almost universally, although unconsciously,
applied in state estimation of one-qubit systems.  For these systems, mutually
unbiased bases consisting of the eigenstates
of the $\sigma_x$, $\sigma_y$ and $\sigma_z$ operators have been
the standard choice for tomographic measurements since the very first
studies in polarimetry\cite{Stokes1852}.  

For systems of qubits MUBs can be constructed as mutual
eigenstates of Pauli operators following the approach of
\cite{Lawrence2002}.  The particular set
used in this work is shown in the right column of table
\ref{Table1}.  Three of the two-qubit MUBs are separable and
two of them are maximally entangled, making them amenable to
standard linear-optics techniques for projective measurements\cite{Weinfurter1994}. 

To study the advantage of MUBs for state
estimation, we repeated the measurements required for tomography over $3000$
$0.2$-second intervals for each basis for both SSQST
and MUB tomography.  During each interval an average of
$28$ photon pairs per basis were detected.  We added together randomly selected
data sets from among these trials to obtain different numbers $N_{\text{tot}}$ of
total counts summed over all bases. At total numbers
above approximately $10^5$, calculation of the
infidelity was limited by computational errors and the infidelity
failed to continue its monotonic decrease.  The plots in figure \ref{data} were
truncated at a point before this limit.  At each value of $N_{\text{tot}}$  
we performed maximum-likelihood fitting to
find the density matrix most likely to have generated the
dataset, and used the infidelity measure to compare it to the density
matrix fit of the entire dataset containing all counts.  This process was repeated 30 times per point and
the infidelity was averaged to produce the plots in figure
\ref{data}.  The error bars represent the measured standard deviation over the 30 trials.  

The experimental apparatus used to perform state tomography both
in the mutually unbiased bases and in the standard separable bases is
shown in figure \ref{apparatus}.  We generate our two-photon
states by spontaneous parametric downconversion (SPDC) in two $\beta$-BBO crystals cut for type-I
phasematching at a $3^\circ$ opening angle\cite{Kwiat1999}.  Because
the crystals had their axes oriented at 90$^\circ$ to each other, the
source could produce states with a degree of entanglement
controlled by the pump polarization.  The crystals were pumped by a 405-nm
diode laser, generating broadband 
SPDC centered at $810$ nm.  The polarizations of the
two downconverted photons could be controlled by liquid crystal
waveplates (LCWPs) and half waveplates (HWPs).  This control, taken together with
the control on the pump polarization, allowed the system to produce a wide variety of entangled and
unentangled pure states.  Due to the multimodal collection system and
the 0.5 nm bandwidth of our pump laser, the purity of maximally entangled
states generated with the system was $0.9$.  By applying random phase shifts with
the LCWPs, mixed states could also be generated\cite{Adamson2007_2}
with fidelity $>0.98$.

Both maximally entangling and separable
measurements are required for MUB tomography.  For the
maximally entangling measurements, a polarization rotation
followed by two-photon interference on a
50-50 beamsplitter\cite{HOM1987} was
used.  

The visibility of the two-photon interference was measured to be
$93\%$.  Simulations demonstrated that this imperfect visibility
increased infidelity for a given number of counts by between
zero and $6\%$, depending on the state.  This imperfect visibility
was taken into account in constructing the operator basis used
in the maximum-likelihood fitting algorithm.  Instead of
consisting of projectors onto pure states, the entangling
measurements were modeled as rank-2 operators equal to a weighted sum of 
a projector onto $\ket{\psi^+}$ and a projector onto $\ket{\psi^-}$.
This change did not affect the informational completeness of the whole
set of measurements, but meant that a greater number of
counts needed to be obtained to correctly estimate the
expectation values of operators that depended on these partially mixed projectors.

The beamsplitter was mounted on a scissor jack and so could be
removed from the optical path without changing the alignment,
allowing us to implement separable measurements as well as
entangling measurements.  Standard
polarization analysis enabled us to collect measurements from
the three separable MUBs and the nine separable bases in
SSQST.

Figure \ref{data} shows plots of the infidelity $\left(1-F\right)$ against total
number of counts $N_{\text{tot}}$ for representative mixed, entangled and
separable states.  Based on previous analyses \cite{deBurgh2008} we
expect the infidelity to drop as $1/\sqrt{N_{\text{tot}}}$ for pure
states and $1/N_{\text{tot}}$ for maximally mixed states.  The state
$\ket{HV}$ and the maximally mixed state agree with this prediction,
but the infidelity of the entangled state
$\frac{1}{\sqrt{2}}\left(\ket{HH}+\ket{VV}\right)$ drops more rapidly
than predicted.  This effect is not currently understood but we are
investigating it in follow-up experiments.  The apparent crossing of the MUBs
and SSQST infidelity lines is believed to be due to numerical errors for
larger $N$.  

The mixed state is analytically and conceptually simple because
it generates a uniform probability
distribution in all bases.  For this state the ratio of SSQST infidelity to MUB QST
infidelity was $1.49\pm0.05$, independent of $N_{\text{tot}}$.  This result is reasonably consistent with
the infidelity estimated analytically from the covariance matrix
calculated from a linear inversion formula\cite{James2001}.  This
analysis predicted a value of 1.38, but required the approximation
that the inversion was linear instead of maximum-likelihood and that
the error in the total counts for each basis was uncorrelated to the
individual counts for each measurement in the basis.  The smaller number of bases in MUB
tomography allows 9/5 more measurements to be made in each for
the same number of total copies of the state.  Although SSQST
provides a more complete covering of the Hilbert space by including a greater number of bases, this
does not make up for the greater number
of copies that MUB tomography can distribute to each of its minimal
number of bases.  

It might also be expected that MUBs offer a better estimate of
entangled states.  When we measured the state
$\frac{1}{\sqrt{2}}\left(\ket{HH}+\ket{VV}\right)$ with the two
techniques we found that the ratio of infidelity observed
with SSQST to that with MUB tomography was $1.84\pm0.06$.  In this case, in
addition to the better statistics obtained by having fewer
bases, the MUB tomography is able to estimate the strength of
correlations in different bases without having to collect
redundant information about the single-qubit polarization.

The advantage
of MUB tomography as compared to SSQST
approaches insignificance when we look at separable states.  For
$\ket{HV}$ we observe that the ratio of infidelity for SSQST
 as compared to MUB QST is $1.09\pm0.05$.  However, even the fact
that MUBs are \emph{no worse} at estimating
separable states is indicative of their superior capabilities since
this is the class of pure states that the
standard separable tomography estimates best.

We have demonstrated optimal projective
quantum state tomography on a number of quantum states and
compared it with standard separable-state tomography.  While
high-quality entangling measurements remain difficult to make in
quantum optical systems, for systems where strong entangling
interactions are available, such as trapped ion quantum
computers, MUB tomography may already be a good choice to
reduce the complexity and the duration of quantum state estimation\cite{Klimov2008}.  MUBs are the
natural choice of tomographic bases because of their ability to
eliminate redundant measurements and to provide the best
estimate of a quantum state from measurements on a discrete
number of copies.  Because they are based on PVMs, they can be implemented relatively easily in
multi-particle systems.  This is in contrast to other theoretically
optimal measurements like SIC-POVMs\cite{Ling2006} and to schemes that
require joint measurements over multiple copies of the state\cite{Massar1995}.

The authors thank Andrei Klimov and Robin Blume-Kohout for helpful discussions and
Robert Kosut for providing maximum-likelihood Matlab code.
This work was supported by the Natural Science and Engineering
Research Council of Canada, QuantumWorks, Ontario Centres of Excellence, the
Canadian Institute for Photonics Innovation and the Canadian
Institute for Advanced Research.
\bibliography{../../paper}
\end{document}